\documentclass[a4paper,12pt]{article}
\newcommand{\bra}[1]{\langle{#1}|}
\newcommand{\bbra}[1]{\langle\langle{#1}|}
\newcommand{\ket}[1]{|{#1}\rangle}
\newcommand{\kket}[1]{|{#1}\rangle\rangle}

\newcommand{\BE}{\begin{equation}}
\newcommand{\EE}{\end{equation}}
\newcommand{\BEA}{\begin{eqnarray}}
\newcommand{\EEA}{\end{eqnarray}}

\newcommand{\ad}{a^\dagger}

\newcommand{\Sd}{S^\dagger}
\newcommand{\bz}{{\bar z}}
\newcommand{\bC}{{\bar C}}
\newcommand{\Cz}{C^{(0)}}
\newcommand{\bCz}{{\bar C}^{(0)}}
\newcommand{\bW}{{\bar W}}
\newcommand{\bT}{{\bar T}}
\newcommand{\bD}{{\bar D}}
\newcommand{\bb}{{\bar b}}
\newcommand{\ssl}{\left[ }
\newcommand{\ssr}{\right] }

\newcommand{\NN}{\nonumber}
\newcommand{\bm}{{\bar m}}
\newcommand{\bn}{{\bar n}}
\newcommand{\B}[1]{\bar{#1}}

\newcommand{\Tr}{{\rm Tr\,}}
\newcommand{\bfT}{{\bf{T}}}

\newcommand{\bfD}{{\bf{D}}}

\newcommand{\bfz}{{\bf 0}}
\newcommand{\bfm}{{\bf m}}
\newcommand{\bfn}{{\bf n}}
\newcommand{\bfe}{{\bf e}}
\newcommand{\EQ}{\begin{equation}}
\newcommand{\EN}{\end{equation}}
\newcommand{\bea}{\begin{eqnarray}}
\newcommand{\ena}{\end{eqnarray}}
\newcommand{\vs}[1]{\vspace{#1 mm}}
\newcommand{\hs}[1]{\hspace{#1 mm}}
\renewcommand{\a}{\alpha}
\renewcommand{\b}{\beta}

\renewcommand{\d}{\delta}
\newcommand{\e}{\epsilon}
\def\bbox{{\,\lower0.9pt\vbox{\hrule \hbox{\vrule height 0.2 cm
\hskip 0.2 cm \vrule height 0.2 cm}\hrule}\,}}
\newcommand{\dsl}{\pa \kern-0.5em /}

\newcommand{\pa}{\partial}
\renewcommand{\t}{\theta}

\newcommand{\nn}{\nonumber\\}
\newcommand{\p}[1]{(\ref{#1})}

\newcommand{\ran}{\rangle}

\newcommand{\wt}{\widetilde}

\begin{document}

\topmargin 0pt
\oddsidemargin 0mm

\renewcommand{\thefootnote}{\fnsymbol{footnote}}
\begin{titlepage}

\setcounter{page}{0}
\begin{flushright}
YITP-01-32\\
OU-HET 381 \\
hep-th/0105079
\end{flushright}

\vs{10}
\begin{center}
{\Large\bf Supersymmetry, Spectrum and Fate of D0-D$p$ Systems with $B$-field}
\vs{15}

{\large
Akira Fujii\footnote{e-mail address: fujii@yukawa.kyoto-u.ac.jp},
Yasuyuki Imaizumi\footnote{e-mail address: imaizumi@het.phys.sci.osaka-u.ac.jp}
and
Nobuyoshi Ohta\footnote{e-mail address: ohta@phys.sci.osaka-u.ac.jp}} \\
\vs{10}
${}^\ast${\em Yukawa Institute for Theoretical Physics, Kyoto University,
Kyoto 606-8502, Japan}

\vs{5}
${}^{\dagger\:\ddagger}${\em Department of Physics, Osaka University,
Toyonaka, Osaka 560-0043, Japan}

\end{center}
\vs{15}

\centerline{{\bf{Abstract}}}
\vs{5}

It has been shown that D0-D$p$ $(p=2, 4, 6, 8)$ systems can be BPS in the
presence of $B$-field even if they are not otherwise. We review the number
of remaining supersymmetries, the open string ground state spectrum and
the construction of the D0-D$p$ systems as solitonic solutions in the
noncommutative super Yang-Mills theory. We derive the complete mass spectrum
of the fluctuations to discuss the stability of the systems. The results
are found to agree with the analysis in the string picture. In particular,
we show that supersymmetry is enhanced in D0-D8 depending on the
$B$-fields and it is consistent with the degeneracy of mass spectrum.
We also derive potentials and discuss their implications for these systems.

\end{titlepage}
\newpage
\renewcommand{\thefootnote}{\arabic{footnote}}
\setcounter{footnote}{0}

\section{Introduction}

D-branes in the background of constant $B$-field have many rich and
interesting structures, one of which is that this system can be described by
noncommutative theory~\cite{CDS,DH}. The relevance of noncommutative theory
was first noticed in the Matrix model context~\cite{HW}, and the
D-branes with $B$-field and noncommutative theories have been studied by
many authors after the appearance of \cite{SW}. An important development in
this direction is the construction of soliton solutions in scalar
noncommutative field theory~\cite{GMS}, and solitons in noncommutative
scalar-gauge theory have been much studied~\cite{HKLM}-\cite{MGJ}.
In particular, a simple technique to construct exact solutions in various
noncommutative gauge theories is introduced in ref.~\cite{HKL}. This
solution generating technique has been used to construct many noncommutative
soliton solutions which represent D-brane bound states~\cite{HKL,LST}.
These solutions turned out to play a significant role in the study of tachyon
condensation~\cite{Sen,HKM}. The stability of these system thus constitutes
an interesting subject.

In the absence of a $B$-field, the D0-D0, D0-D4 and D0-D8 systems in type
IIA superstrings are supersymmetric (BPS), while the others are not.
It has been discovered that the BPS conditions for the D0-D$p$ $(p=2,4,6,8)$
systems are modified and those hitherto considered non-BPS can be BPS in the
presence of the $B$-field~\cite{W,BBH}. This can be checked by evaluating the
ground state energy~\cite{CIMM} or by examining Killing
spinors~\cite{MPT,OTom,OTown}. The BPS D0-D4, for example, remains so only
when the $B$-field is (anti-)self-dual and non-BPS D0-D6 becomes BPS
for appropriate $B$. A new supersymmetric locus has also been found for a
certain background for mysterious D0-D8 system. Stability around these
BPS conditions was also partially analyzed in the string picture.

In view of these developments, it is interesting to explore the soliton
realization of these brane systems and study their properties. In particular,
we are interested in how the new supersymmetric locus mentioned above is
different from others. The stability of the D0-D$p$ system observed in the
string picture should also be reproduced in the framework of noncommutative
theories. In this paper, we construct the noncommutative soliton solutions
corresponding to the D0-D$p$ systems and consider the small fluctuations
around the solutions. The solutions for D0-D2 and D0-D4 have been
discussed in ref.~\cite{AGMS}. Here we extend the analysis to all the D0-D$p$
systems, obtain the complete mass spectrum of the fluctuations and find
that it is consistent with supersymmetry in the string consideration
including its degeneracy. The mass spectrum also gives the precise stability
conditions of the brane systems, which also agree with those obtained in
the string picture. Finally we derive the potentials for possible tachyonic
scalar modes and discuss the fate of the brane systems.

This paper is organized as follows. In sect.~2.1, we describe the BPS
conditions of the D0-D$p$ system with $B$-field and the form of the conditions
in the zero slope limit. In sect.~2.2, we also analyze the zero point energy
of the D0-D$p$ system~\cite{CIMM,W}, and then discuss the stability of this
system in comparison with the BPS conditions obtained in sect.~2.1.
In sect.~3, we use noncommutative super Yang-Mills theory on the D$p$-branes
with $B$-field to realize the D0-D$p$ system as a noncommutative soliton in the
theory~\cite{GMS,HKLM,AGMS,HKL}. We then derive mass spectrum in the small
fluctuations of this soliton solution. In the process the gauge mode in
the fluctuations is identified with the help of the Gauss law.
Details of the calculations are summarized in the appendix.
We show that the tachyonic modes are present (absent) when instability
(stability) is obtained in the string picture analysis in sect.~2.
In one supersymmetric locus of the D0-D8 system, we show that there is no
massless spectrum but the degeneracy of the spectrum correctly reflects
the enhancement of supersymmetry of this system. In the other new
supersymmetric locus, there is massless spectrum but we find that no
enhancement of supersymmetry is possible. Finally in sect.~4, we derive
potentials for the system.

\section{D0-D$p$ with a Constant $B$-field in the String Picture}
\label{sec:2}

Let us consider the D$0$-D$p$ ($p=2,4,6,8$) systems in Type IIA theory with a
constant $B$-field. The D$p$-brane fills the directions $x_0,\cdots,x_{p}$
and the $B$-field is block-diagonal and lives in the directions
$(x_1,x_2,\cdots,x_{p-1},x_p)$:
\begin{equation}
B = \mbox{diag}([B_1],\cdots,[B_{p/2}])=
\frac{\e}{2\pi\alpha^\prime}\mbox{diag}([b_1],\cdots,[b_{p/2}]),
\label{bfield}
\end{equation}
where $[B_i]$ and $[b_i]$ $(i=1,\cdots,p/2)$ are $2\times2$ matrices
\begin{equation}
[B_i] =
\left(\begin{array}{cc}
0 & -B_i \\
B_i & 0
\end{array}\right)
=\frac{\e}{2\pi\alpha^\prime}[b_i]=\frac{\e}{2\pi\alpha^\prime}
\left(\begin{array}{cc}
0 & -b_i \\
b_i & 0
\end{array}\right).
\label{eq:b-field}
\end{equation}
The metric on the string worldsheet is written as $g_{ab}=\e \d_{ab}\;
(a,b=1,\cdots,p), g_{00}=-1$. Here $\e$ is a parameter to define the zero
slope limit so as to give noncommutative theories~\cite{SW,CIMM}:
\begin{equation}
\alpha^\prime \sim \e^{1/2}\to 0, \quad
B:\mbox{ finite}, \quad
b_i \sim \e^{-1/2} \to \infty.
\label{eq:zeroslope}
\end{equation}

In this section we summarize the results on the BPS conditions of
D$0$-D$p$ with the $B$-field and the ground state energies~\cite{W,CIMM}.
The BPS conditions can be obtained by examining the constraints on the
Killing spinor~\cite{OTom,OTown}. In D$0$-D$8$, the supersymmetry is
enlarged when the $B$-field satisfies a certain condition~\cite{OK,OTown}.

\subsection{BPS condition}
\label{sec:2.1}

Type IIA theory has 32 supercharges of opposite chirality, $Q_\alpha$ and
$\wt{Q}^\beta$ that originate from the left and right moving modes on the
string worldsheet, respectively. In the presence of a D-brane, some linear
combinations of the supersymmetries $\sum_\a \e^\a Q_\a + \sum_\b \wt{\e}_\b
\wt{Q}^\b$ remain unbroken while the others become broken.

Putting
\begin{equation}
\tan \pi v_i = b_i,\;\;\; -\frac{1}{2}<v_i<\frac{1}{2},
\label{eq:b2v}
\end{equation}
we find that the BPS condition for unbroken supersymmetry is
\begin{equation}
\epsilon_\beta =
(\Gamma^1\cdots\Gamma^{p}e^{\pi v_1 \Gamma^{12}}\cdots
e^{\pi v_{p/2} \Gamma^{p-1,p}})_{\b\a}\e^\a,
\label{eq:bps}
\end{equation}
and $\wt\e$ of opposite chirality is determined in terms of $\e$.
If (\ref{eq:bps}) has a nontrivial solution, the corresponding supersymmetry
remains unbroken. In fact, it happens when $v_1,\cdots,v_{p/2}$ in \p{eq:b2v}
satisfy certain conditions, which are summarized in what follows.

\noindent
$\bullet$ \underline{\it D0-D6}\\
For $p=6$, the BPS condition is determined from \p{eq:bps} as
\begin{equation}
\pm v_1\pm v_2\pm v_3 = \pm\frac{1}{2},
\label{eq:bpsd6}
\end{equation}
where the combination of the three $\pm$ signatures is arbitrary.
By appropriate reversal of the coordinate axes, we can always arrange so that
$v_1$, $v_2$, and $v_3$ (and hence $b_i$) are non-negative.
For non-negative $(v_1, v_2, v_3)$ satisfying (\ref{eq:bpsd6}),
there are {\it four} unbroken supercharges (referred to as $1/8$ SUSY).

All the BPS conditions in \p{eq:bpsd6} are not independent; independent
relations are\footnote{
In Witten's convention~\cite{W}, the signs of $v_i$'s are arbitrary and
the conditions can be written simply as \p{eq:d6+++}. In our convention
with $v_i\geq 0$, the two cases \p{eq:d6+++} and \p{eq:d6-++} must be
distinguished. Similar remarks apply to other cases.}
\begin{eqnarray}
v_1+v_2+v_3&=&\frac{1}{2},  \label{eq:d6+++}\\
-v_1+v_2+v_3&=&\frac{1}{2}, \;\;\; (\mbox{and its permutations})
\label{eq:d6-++}.
\end{eqnarray}

It is convenient to parametrize these BPS conditions by the four parameters
$r_0=\frac12-(v_1+v_2+v_3)$ and $r_i=r_0 +2v_i\,\,(i=1,2,3)$.
In the zero slope limit (\ref{eq:zeroslope}), since $v_i$ goes like
\begin{equation}
v_i \to \frac{1}{2}\mbox{sign}\:b_i-\frac{1}{\pi b_i},
\end{equation}
the parameter $r_0$ behaves as
\begin{equation}
r_0\sim -1+\frac{1}{\pi}
\biggl(\frac{1}{b_1}+\frac{1}{b_2}+\frac{1}{b_3}\biggr).
\end{equation}
We see that the BPS condition $r_0=0$ for \p{eq:d6+++} is not realized
in the zero slope limit~\p{eq:zeroslope}.

On the other hand, $r_i\,\,(i=1,2,3)$ tends to
\begin{equation}
r_i\sim \frac{1}{\pi}
\biggl(\frac{1}{b_1}+\frac{1}{b_2}+\frac{1}{b_3}\biggr)
-{2\over \pi b_i},
\end{equation}
which is consistent with the BPS conditions \p{eq:d6-++} in the zero slope
limit.

\noindent
$\bullet$ \underline{\it D0-D2}\\
For $p=2$, the BPS condition for the unique parameter $v_1$ is
\begin{equation}
v_1\rightarrow\frac{1}{2},
\end{equation}
which is nothing but $b_1 =\infty$, and this is consistent with the zero
slope limit. When this is satisfied, the system possesses $16$ unbroken
supercharges ($1/2$ SUSY).
At this point the system is essentially the D$0$-D$0$ system.

\noindent
$\bullet$ \underline{\it D0-D4}\\
For $p=4$, the independent BPS condition is
\begin{equation}
v_1-v_2=0.
\label{eq:d4bps}
\end{equation}
When this is satisfied, the system has {\it eight} unbroken
supercharges ($1/4$ SUSY). Near the zero slope limit, the BPS condition
(\ref{eq:d4bps}) reduces to
\begin{equation}
\frac{1}{b_1}-\frac{1}{b_2}=0.
\label{eq:d4bpsb}
\end{equation}

\noindent
$\bullet$ \underline{\it D0-D8}\\
For $p=8$, the BPS condition is
\begin{equation}
\pm v_1\pm v_2\pm v_3\pm v_4 = {\rm integer}.
\end{equation} With the assumption $v_i\geq 0,\,(i=1,2,3,4)$,
its independent branches are
\begin{eqnarray}
v_1+v_2+v_3+v_4&=&1, \label{eq:d8++++1}\\
-v_1+v_2+v_3+v_4&=&1 \;\;\;\mbox{(and its permutations)},
\label{eq:d8-+++1}
\end{eqnarray}
and
\begin{eqnarray}
-v_1-v_2+v_3+v_4&=&0 \;\;\;\mbox{(and its permutations)},
\label{eq:d8--++0}\\
-v_1+v_2+v_3+v_4&=&0 \;\;\;\mbox{(and its permutations)}
\label{eq:d8-+++0}.
\end{eqnarray}
The cases \p{eq:d8++++1} and \p{eq:d8-+++1} are the new loci of supersymmetry
found in ref.~\cite{W}, and it is interesting to examine how the theory
looks like there.

When one of the conditions (\ref{eq:d8++++1})-(\ref{eq:d8-+++0}) is satisfied,
the system has {\it two} unbroken supercharges ($1/16$ SUSY).
For the third case \p{eq:d8--++0}, say $-v_1-v_2+v_3+v_4=0$, supersymmetry
may be enhanced if additional conditions are satisfied. If
\begin{equation}
v_1=v_3\neq v_2=v_4, \quad {\rm or} \quad
v_1=v_4\neq v_2=v_3,
\label{1/8susy}
\end{equation}
is obeyed, the number of the unbroken supercharges is raised to
{\it four} ($1/8$ SUSY). If further
\begin{equation}
v_1=v_2=v_3=v_4,
\label{3/16susy}
\end{equation}
{\it six} of the supersymmetries remain unbroken ($3/16$ SUSY)~\cite{OK,OTown}.

There is no case that supersymmetry is enhanced for the new loci~\p{eq:d8++++1}
and \p{eq:d8-+++1}.

In the zero slope limit ($v_i \to \frac12$), only (\ref{eq:d8-+++1}) and
(\ref{eq:d8--++0}) can be realized. These conditions may be parametrized by
$s_i =1-(v_1+v_2+v_3+v_4)+2v_i\,(i=1,2,3,4)$ and
$t_{ij}=-(v_1+v_2+v_3+v_4)+2v_i +2v_j \,(i,j=1,\cdots,4;\, i>j)$.
Near the zero slope limit, they tend to
\begin{equation}
s_1\sim \frac{1}{\pi}\biggl(-\frac{1}{b_1}+\frac{1}{b_2}
+\frac{1}{b_3}+\frac{1}{b_4}\biggr),\quad
t_{12}\sim \frac{1}{\pi}\biggl(-\frac{1}{b_1}-\frac{1}{b_2}
+\frac{1}{b_3}+\frac{1}{b_4}\biggr).
\label{d0d8bps}
\end{equation}
These parameters are used in the following analysis.

\subsection{Stability of the ground states}
\label{sec:2.2}

In this subsection we summarize the ground state energies of D$0$-D$p$
from the string and discuss the stability of the systems~\cite{AGMS,W}.
To compute the ground state energies, we take the light-cone gauge as usual.
Since we know that the zero point energies in the Ramond sector always
vanish owing to the worldsheet supersymmetry, we only need to pay attention
to the transverse coordinates $X^1,\cdots,X^8$ and their superpartners
$\psi^1_{\pm},\cdots,\psi^8_{\pm}$ in the Neveu-Schwarz sector.
It is practical to utilize the complex coordinates and spinors,
$X_i= X^{2i-1}+iX^{2i}$, $\bar{X}_i= X^{2i-1}-iX^{2i}$, $\psi_{i}
= \psi^{2i-1}+i\psi^{2i}$ and $\bar{\psi}_{i}= \psi^{2i-1}-i\psi^{2i}$.

{}From the mode expansion, the complex bosons $X_i$ and fermions $\psi_{i}$
$(i=1,\cdots,p/2)$ give zero point energy
\begin{eqnarray}
\mbox{(boson)}:\hs{3}
\frac{1}{24}-\frac{v_i^2}{2}, \hs{10}
\mbox{(fermion)}: \hs{3}
-\frac{1}{24}+\frac{1}{2}\Bigm(\frac{1}{2}-v_i\Bigm)^2.
\label{eq:casimirp}
\end{eqnarray}
The zero point energy for $X_j$ and $\psi_{j}\; (j= p/2+1,\cdots,4)$
satisfying the Dirichlet boundary conditions is obtained by substituting
1/2 to $v_i$ in (\ref{eq:casimirp}). The total zero point energy of the NS
sector is thus
\begin{equation}
E_0=\frac{p-4}{8}-\frac{1}{2}\sum_{i=1}^{p/2} v_i.
\label{eq:0pt}
\end{equation}
Let us summarize the stability/instability of the D$0$-D$p$ system on the
basis of (\ref{eq:0pt}).

\noindent
$\bullet$ \underline{\it D0-D2}\\
The zero point energy is $E_0=-\frac12(\frac12+v_1)$. The BPS condition $v_1
=\frac12$ gives $E_0=-\frac12$. Because $\psi_{i}$ and $\bar{\psi}_{i}$
have mode expansions
\begin{equation}
\psi_{i}=\frac{1}{\sqrt{2}}\sum_{n=-\infty}^\infty
e^{i(n+v_i)(\tau-\sigma)}\psi_{i,n+v_i},\quad
\bar{\psi}_{i}=\frac{1}{\sqrt{2}}\sum_{n=-\infty}^\infty
e^{i(n-v_i)(\tau-\sigma)}\bar{\psi}_{i,n-v_i},
\end{equation}
$\psi^\dagger_{1,v_1}\ket{E_0 }$ is massless, where
$\psi^{\dagger}_{i,v}=\bar{\psi}_{i,-v}$.
Thus, in this supersymmetric case, a tower of massive states stands on this
massless state after the GSO projection is applied.

On the other hand, when this system is not BPS ({\it i.e.} $0\leq v_1< 1/2$),
the state $\psi^\dagger_{1,v_1}|E_0\ran$ has a negative energy
$E_1 =-\frac{1}{2}(\frac{1}{2}-v_1)$, which means that the system is unstable.

\noindent
$\bullet$ \underline{\it D0-D4}\\
The zero point energy is $E_0=-\frac12(v_1+v_2)$.
If the BPS condition (\ref{eq:d4bps}) is satisfied, we see $E_0=-v_i$,
which is negative. However, just as in the D0-D2 case, the ground state is
the massless state $\psi^\dagger_{i,v_i}|E_0 \ran$ chosen by the GSO
projection.

If the BPS condition is not realized, the energy of the first excited state
$\psi_{2,v_2}^\dagger |E_0\ran$ is $\frac12(v_2-v_1)$ and
that of $\psi_{1,v_1}^\dagger |E_0\ran$ is $\frac12(v_1-v_2)$.
Thus, the D0-D4 system always contains a tachyonic mode unless $v_1-v_2=0$.

\noindent
$\bullet$ \underline{\it D0-D6}\\
The zero point energy is $E_0=\frac14-\frac12(v_1+v_2+v_3)$.
The BPS condition is (\ref{eq:d6+++}) or (\ref{eq:d6-++}).

Let us take the BPS conditions (\ref{eq:d6-++}), say,
$-v_1 +v_2 +v_3=\frac12$, and we have $E_0=-v_1$. The first excited state
$\psi^\dagger_{1,v_1}|E_0 \ran$ is massless which is kept by the GSO
projection. Because each state $\psi^\dagger_{i,v_i}|E_0\ran \;(i=1,2,3)$
has energy $E_i=r_i/2$, the system contains a tachyonic mode when one of
the $r_i$ is negative, while the system is stable if all $r_i$ are positive.
Notice that there is no case that two of $r_i$ are simultaneously negative.

On the other hand, for the BPS condition (\ref{eq:d6+++}), we get $E_0=0$.
Since the R sector contains a massless state, this massless state $|E_0\ran$
in the NS sector should be kept by the GSO projection. The system is stable
(unstable) if $r_0 $ is positive (negative).

\noindent
$\bullet$ \underline{\it D0-D8}\\
The zero point energy is $E_0=\frac{1}{2}[1-(v_1+v_2+v_3+v_4)]$.
D$0$-D$8$ has, up to the permutations of $\pm$ signatures, four branches
of the BPS conditions, (\ref{eq:d8++++1})-(\ref{eq:d8-+++0}).

First, at the new BPS locus (\ref{eq:d8++++1}), the system contains a unique
massless state $|E_0\ran$. Close to this point, the system is stable (unstable)
if $1-(v_1+v_2+v_3+v_4)>0$ ($<0$).

Second, at another new BPS locus (\ref{eq:d8-+++1}), the stability of the
system is the same as that of D$0$-D$6$. The system is unstable when one of
the $s_i$ defined above \p{d0d8bps} is negative, and is stable when all of
$s_i$ are positive. There is no case in which two of $s_i$ are simultaneously
negative. There is no enhancement of supersymmetry at these two loci.

Third, for the BPS conditions (\ref{eq:d8--++0}), we find $E_0=\frac12-v_i-
v_j$. This energy $E_0$ can be negative if $v_i$ and $v_j$ are sufficiently
large. Thus, we should keep the four states $\psi_{k,v_k}^\dagger|E_0\ran \;
(k=1,\cdots,4)$ with energies $\frac12-v_i \; (i=1,\cdots,4)$ in the GSO
projection. Since all these states have positive energies, the system is
always stable for these BPS conditions. Though there is no massless state at
this locus, supersymmetry is enhanced when additional conditions are
satisfied and this is reflected in the mass spectrum. When \p{1/8susy} is
satisfied, the spectrum is doubly degenerate (1/8 SUSY). If \p{3/16susy} is
satisfied, the degeneracy is fourfold, reflecting 3/16 SUSY.

Finally, assuming the BPS conditions (\ref{eq:d8-+++0}), for example,
$-v_1+v_2+v_3+v_4=0$, we obtain $E_0=\frac12-v_1$, which is positive. The state
$|E_0\ran$ is kept by the GSO projection. (In particular, for $v_1\to \frac12$,
the BPS condition reduces to $v_2+v_3+v_4=\frac12$ and the system becomes
D$0$-D$6$, where we should keep the massless state $|E_0\ran$.)
Since $|E_0\ran$ has a positive energy at this BPS locus,
this system is stable around this BPS condition.

\section{D$0$-D$p$ Solution and Mass Spectrum in the Solitonic Realization}
\label{sec:3}
\setcounter{equation}{0}

In this section we concentrate on the zero slope limit (\ref{eq:zeroslope}),
and consider the corresponding $(p+1)$-dimensional noncommutative $U(1)$ gauge
theory~\cite{CDS,DH,SW}
\BE
S = -{1\over 4g_{\rm YM}^2 G_s/g_s}
\int\,dt\,d^{p}x\,\sqrt{-G} G^{\mu\lambda} G^{\nu\sigma}
 F_{\mu\nu} * F_{\lambda\sigma},
\label{eq:YM}
\EE
where $g_s$ is the string coupling, $g^2_{\rm YM}=(2\pi)^{p-2}(\a')^{(p-3)/2}
 g_s$ and
\bea
G_{ab} = g_{ab} - (2\pi\a')^2 (Bg^{-1}B)_{ab}
&\to& \e b^2 \d_{ab}, \nn
G_s = g_s \left(\frac{\mbox{det}(g+ 2\pi\a' B)}{\mbox{det}\, g}\right)
^{\frac{1}{2}} &\to& g_s \prod_{i=1}^{p/2} b_i,
\ena
in the zero slope limit. Though we should supplement \p{eq:YM} with
fermionic terms when some supersymmetry is preserved, it is enough to
consider only the bosonic terms for our purpose.

The noncommutativity of the space is manifested in the relation
\BE
\ssl x^{2i-1},x^{2i} \ssr =i\theta_i,\quad
\theta_i = {2\pi \a' \over \e b_i} = \frac{1}{B_i}, \quad
(i=1,\cdots,p/2),
\EE
where we assume $b_i, \t_i \geq 0$ as in sect.~\ref{sec:2}.
Let us define complex coordinates
\BE
z_{j}={1\over\sqrt{2}}(x^{2j-1}+ix^{2j}),\quad
\bz_{j}={1\over\sqrt{2}}(x^{2j-1}-ix^{2j}),
\EE
and creation/annihilation operators $\ad_i =\bz_i /\sqrt{\theta_i}$ and
$a_i = z_i /\sqrt{\theta_i}$. Thus we use the operator formalism for the
noncommutative theories. In the temporal $A_0 =0$ gauge,
we can rewrite (\ref{eq:YM}) as
\BEA
S = -{\prod_{i=1}^{p/2} (2\pi \bar b_i) \over g^2_{\rm NYM}}
\int&\!\!\!\!dt\!\!\!\!&{\rm Tr}\left[ \sum_{i=1}^{p/2}
\left(
-\pa_{t}C_i\pa_{t}\bC_i +\frac{1}{2}\left( \ssl C_i , \bC_i\ssr
+\frac{1}{\bb_i}\right)^2\right)\right.\nn
&&\left. +\sum_{i<j} \left( \ssl C_i , \bC_j\ssr\ssl C_j , \bC_i\ssr
+\ssl C_i , C_j\ssr\ssl \bC_j , \bC_i\ssr \right)\right],
\label{action}
\EEA
where we have set $g^2_{\rm NYM} = g^2_{\rm YM}\prod_{i=1}^{p/2} b_i$,
$\bar b_i=\e b_i^2 \t_i = 2\pi\a' b_i$ and
\BE
C_j = C_{z_j}=\frac{1}{\sqrt{\e}b_j}\Big(-iA_{z_j}+\frac{1}{\sqrt{\t_j}}
\ad_j\Big), \quad
\bC_j = C_j^{\dagger}=\bC_{\bz_j}=\frac{1}{\sqrt{\e}b_j}\Big(iA_{\bz_j}
+\frac{1}{\sqrt{\t_j}}a_j \Big).
\EE
In addition to the equations of motion, the gauge condition $A_0 =0$ induces
the Gauss law constraint
\begin{equation}
\label{Gauss}
\sum_{i=1}^{p/2} \left( \ssl C_i , \pa_t\bC_i\ssr
 + \ssl \bC_i, \pa_t C_i\ssr \right) =0.
\end{equation}

On D$0$-D$p$, we can construct exact solitonic solutions of (\ref{action})
by applying solution generating technique~\cite{HKL,AGMS}.
This technique enables us to find new solutions, which generally have nonzero
soliton numbers, from a trivial solution by an almost gauge transformation
with shift operator.

To define the shift operator $S$, we prepare an ordering of the states
\BE
\ket{n_1,\cdots,n_{p/2}}=
\prod_{i=1}^{p/2}{1\over\sqrt{n_i !}}(\ad_i)^{n_i}\ket{0,\cdots,0}.
\EE
Two sets of $p/2$ non-negative integers, ${\bf m}=(m_1,\cdots,m_{p/2})$
and ${\bf n}=(n_1,\cdots,n_{p/2})$, for which we define
$\bm_j=\displaystyle{\sum_{i=j}^{p/2}}m_i$ and
$\bn_j=\displaystyle{\sum_{i=j}^{p/2}}n_i$ with $j=1,\cdots,p/2$,
are ordered by the following rules:
\begin{enumerate}
\item If $\bm_j=\bn_j$ for all $1\leq j\leq \frac{p}{2}$,
${\bf m}={\bf n}$.
\item If $\bm_j=\bn_j\; (j=1,\cdots,k-1)$ and
$\bm_k>\bn_k$ for some
$k$ ($1\leq k\leq \frac{p}{2}$), ${\bf m}>{\bf n}$.
\item If $\bm_j=\bn_j\; (j=1,\cdots,k-1)$ and
$\bm_k<\bn_k$ for some
$k$ ($1\leq k\leq \frac{p}{2}$), ${\bf m}<{\bf n}$.
\end{enumerate}
We can order all the states by this rule. For example, in D$0$-D$4$,
this orders the states as
\BEA
&&\kket{0}=\ket{0,0},\NN\\
&&\kket{1}=\ket{1,0},\quad\kket{2}=\ket{0,1},\NN\\
&&\kket{3}=\ket{2,0},\quad\kket{4}=\ket{1,1},\quad
\kket{5}=\ket{0,2},\cdots,\NN\\
&&\kket{
\scriptscriptstyle{{1\over 2}(n_1 +n_2 )^2
+{1\over 2}(n_1 +3n_2)}}
=\ket{n_1,n_2},\cdots.
\EEA
With these preparations, we define the shift operator
\BE
S=\sum_{i=0}^{\infty}\kket{i}\bbra{i+1},
\label{eq:shift}
\EE
which annihilates $\ket{0,\cdots,0}$ and satisfies
\BE
S\Sd =1,\quad \Sd S=1-P_0,
\EE
where $P_0 =\ket{0,\cdots,0}\bra{0,\cdots,0}$ is a projection operator onto
the vacuum.

As the trivial solution, we take
\begin{equation}
\label{sol}
\Cz_i=\frac{1}{\sqrt{\bar b_i}} \ad_i,
\end{equation}
which satisfies both the equation of motion and the Gauss law constraint.
Using the shift operator (\ref{eq:shift}), we can obtain another simple
solution
\BE
\Cz_j = \frac{1}{\sqrt{\bar b_j}} \Sd\ad_j S.
\label{sol2}
\EE
It is also possible to use an arbitrary power of $S$ to generate new
solutions, but this does not essentially change the following results.

Let us investigate small fluctuations around the exact solution (\ref{sol2})
represented by
\begin{eqnarray}
\hs{-5}C_i\!\!&=&\!\!\Cz_i+\delta C_i \nn
   \!\!&=&\!\!\Cz_i+P_0A_iP_0+P_0W_i(1-P_0)+(1-P_0)\bT_iP_0+
S^\dagger D_iS.
\label{fluc}
\end{eqnarray}
The mass matrices of the fluctuations obtained by substituting (\ref{fluc})
to (\ref{action}), and the complete list of the eigenvalues are relegated to
the appendix. In what follows we discuss only the results and compare
the mass spectrum with the analysis of the D0-D$p$ in terms of string in
sect.~\ref{sec:2.2}. In this comparison, we need, not the whole eigenvalues
but only those of the lowest modes, and we examine when they become positive
or negative.  We also show that the degeneracy of the mass spectrum
is in perfect agreement with the number of unbroken supersymmetries in D0-D$p$,
as discussed in sect.~2.

\noindent
$\bullet$ \underline{\it D0-D2}\\
In this case~\cite{AGMS}, the lowest eigenvalue that we have to examine
is that of the matrix element $\bra{0}T_1\Sd\ket{0}$. The value turns out
to be
\begin{equation}
\label{D2mode}
-\frac{1}{\bb_1},
\end{equation}
which is always negative and indicates that the system is always unstable.

\noindent
$\bullet$ \underline{\it D0-D4}\\
The eigenvalues to be examined here are
\begin{eqnarray}
-\frac{1}{\bb_1}+\frac{1}{\bb_2}\;\;\;,\;\;\;
\frac{1}{\bb_1}-\frac{1}{\bb_2},\hs{29}\label{D4mode1}
\end{eqnarray}
one of which is positive and the other is negative unless $1/\bb_1-1/\bb_2=0$.
Therefore this system is unstable except for the case where (\ref{eq:d4bpsb})
is realized.\footnote{Note that $\bb_i$ and $b_i$ are proportional to each
other, so that the BPS conditions in sect.~\ref{sec:2.1} are directly
translated into $\bb_i$.}
In this BPS case, other modes are all massive and we are left
with two massless states from (\ref{D4mode1}). This agrees with the fact
that the system has {\it eight} unbroken supercharges.

\noindent
$\bullet$ \underline{\it D0-D6}\\
The lowest eigenvalues to be considered are
\begin{eqnarray}
-\frac{1}{\bb_1}+\frac{1}{\bb_2}+\frac{1}{\bb_3}\;,\;\;
\frac{1}{\bb_1}-\frac{1}{\bb_2}+\frac{1}{\bb_3}\;,\;\;
\frac{1}{\bb_1}+\frac{1}{\bb_2}-\frac{1}{\bb_3}.
\label{D6mode}
\end{eqnarray}

When one of the BPS conditions (\ref{eq:d6-++}), say, $r_1=0$ is satisfied,
(\ref{D6mode}) becomes all non-negative. The (bosonic) mass spectrum
contains only one massless mode in (\ref{D6mode}). This again agrees
with the fact that this system has {\it four} supercharges in BPS case.

Around the region $r_1=0$, the mass eigenvalues in (\ref{D6mode}) are all
positive if $r_1>0$ and this indicates that the system is stable (other
eigenvalues are all more positive), whereas one of the mass eigenvalues in
(\ref{D6mode}) becomes negative if $r_1 <0$.\footnote{It is noted in
ref.~\cite{W} that the above field theoretical argument is reliable only
in the small $r_i$ region. The analysis for large $r_i$ is beyond the scope
of this paper.} It is important that all the eigenvalues in \p{D6mode},
not just one of them, must be positive for stability.
Thus the analysis of stability/instability of D$0$-D$6$ based on the
noncommutative field theory shows nice agreement with that in the string
analysis in sect.~\ref{sec:2.2}.

\noindent
$\bullet$ \underline{\it D0-D8}\\
The mass eigenvalues to be examined are
\begin{equation}
-\frac{1}{\bb_1}+\frac{1}{\bb_2}+\frac{1}{\bb_3}+\frac{1}{\bb_4}
\;,\;\;
\frac{1}{\bb_1}-\frac{1}{\bb_2}+\frac{1}{\bb_3}+\frac{1}{\bb_4}
\;,\;\;
\frac{1}{\bb_1}+\frac{1}{\bb_2}-\frac{1}{\bb_3}+\frac{1}{\bb_4}
\;,\;\;
\frac{1}{\bb_1}+\frac{1}{\bb_2}+\frac{1}{\bb_3}-\frac{1}{\bb_4}.
\label{D8mode}
\end{equation}
When one of the new BPS conditions (\ref{eq:d8-+++1}), say, $s_1$=0 is
satisfied, (\ref{D8mode}) gives
\begin{equation}
\label{D8mode2}
0\;,\;2\Bigl(\frac{1}{\bb_3}+\frac{1}{\bb_4}\Bigr)\;,\;
2\Bigl(\frac{1}{\bb_2}+\frac{1}{\bb_4}\Bigr)\;,\;
2\Bigl(\frac{1}{\bb_2}+\frac{1}{\bb_3}\Bigr),
\end{equation}
and one massless mode appears. This is consistent with the observation
that D$0$-D$8$ contains {\it two} unbroken supercharges for $s_i=0$.
We note that here is one massless mode but we get no enhancement of
supersymmetry. Similarly to D0-D6, we find that, close to $s_1=0$,
the stability is determined by the sign of $s_1$ since other modes are
all massive.

If one of the BPS conditions (\ref{eq:d8--++0}), for example, $t_{12}=0$ is
satisfied, (\ref{D8mode}) reduces to
\begin{equation}
\label{D8mode3}
\frac{2}{\bb_2}\;,\;\frac{2}{\bb_1}\;,\;
\frac{2}{\bb_4}\;,\;\frac{2}{\bb_3},
\end{equation}
which are all positive and thus the system is stable around $t_{12}=0$.
Moreover, if $b_i$ satisfy $b_1=b_3\neq b_2=b_4$ (or $b_1=b_4\neq b_2=b_3$)
besides $t_{12}=0$, (\ref{D8mode3}) contains two doubly degenerate
eigenvalues, $2/\bb_1$ and $2/\bb_2$, consistent with the fact that the
system has {\it four} unbroken supercharges. If $b_i$ further satisfy the
condition $b_1=b_2=b_3=b_4$, (\ref{D8mode3}) indicates a fourfold degeneracy
in the massive eigenvalues. This is due to the fact that the system has
{\it six} supercharges.

All these results are in agreement with those from the string analysis in
sect.~2.

\section{Scalar Potential from the Yang-Mills Action}
\setcounter{equation}{0}

In this section we evaluate the potentials in (\ref{action}) including the
quartic terms of the fluctuations. They enable us to discuss the ultimate
fate of the brane system when the D0-D$p$ systems are unstable.
We use the notations in sect.~\ref{sec:3}.

The quantity that we evaluate here is
\BEA
V(\bfT,\bfD)=
\sum_{i=1}^{p/2}V_{i\B{i}}(T_i ,D_i )
+\sum_{i<j}V_{ij}(T_i ,T_j ,D_i , D_j )
+\sum_{i<j}V_{i\B{j}}(T_i , T_j ) 
\label{eq:tachyonpotential}
\EEA
where we decompose the sum of the squares of the field strengths as
\BEA
&& \hs{-10} V_{i\B{i}}(T_i ,D_i ) = \Tr
\left(\ssl C_i,\bC_i\ssr +{1\over \bb_i}\right)^2,\NN\\
&& \hs{-10} V_{ij}(T_i ,T_j ,D_i , D_j )=2\, \Tr
\ssl C_i,C_j\ssr\ssl\bC_j,\bC_i\ssr, \quad
V_{i\B{j}}(T_i , T_j )=  2\, \Tr
\ssl C_i,\bC_j\ssr\ssl C_j,\bC_i\ssr.
\label{eq:tpot}
\EEA

In the action (\ref{action}) the field $A_i$ does not have the kinetic
term. From the results in the appendix, all the elements of $T_i$ except
for $\bra{\bfz}T_i\Sd\ket{\bfz}$ and $W_i$ have positive mass eigenvalues.
Therefore, we are allowed to put $A_i = W_i =0$ and
$\bra{\bfz}T_i\Sd\ket{\bfn}=0$ unless $\bfn=\bfz$ in (\ref{eq:tpot}). We then
set $T_i = \bra{\bfz}T_i\Sd\ket{\bfz}$. Furthermore, for simplicity, we assume
\BEA
\label{eq:ansaetze}
\bra{\bfm}D_i \ket{\bfn}&=&0\quad{\rm unless}\quad
m_j = n_j +\delta_{ij},\NN\\
\ssl D_i , D_j \ssr &=& \ssl D_i , \B{D}_j \ssr =
\ssl \B{D}_i , \B{D}_j \ssr = 0\quad{\rm if}\quad i\neq j,\\
\ssl D_i , a_j \ssr &=& \ssl D_i , \ad_j \ssr =
\ssl \B{D}_i , \ad_j \ssr = 0\quad{\rm if}\quad i\neq j.\NN
\EEA

The evaluation $V_{i\B{i}}(T_i ,D_i )$ is already done in~\cite{AGMS}.
The result is
\BEA
V_{i\B{i}}(T_i ,D_i )&=& \left(|T_i|^2 -{1\over\bb_i}\right)^2
+\left(|T_i|^2 - \left|\bra{\bfe_i}D_i\ket{\bfz}+{1\over\sqrt{\bb_i}}\right|^2
+{1\over\bb_i}\right)^2 \NN\\
&& \hs{-10} +\sum_{\bfn\neq\bfz}
\left( \left|
\bra{\bfn}D_i \ket{\bfn-\bfe_i }+\sqrt{n_i\over\bb_i}\right|^2
-\left|\bra{\bfn+\bfe_i }D_i \ket{\bfn}
+\sqrt{n_i +1\over\bb_i}\right|^2 +{1\over\bb_i} \right)^2 ,
\label{eq:vibi}
\EEA
where we should put $\bra{\bfn}D_i\ket{\bfn-\bfe_i}=0$ if
$n_i=0$.
We can also calculate the cross terms $V_{ij}$ and $V_{i\B{j}}$ after
imposing (\ref{eq:ansaetze}) to obtain
\BE
V_{ij}(T_i ,T_j , D_i , D_j )=
2\left| \bra{\bfe_j }D_j \ket{\bfz}+{1\over\sqrt{\bb_j}}
\right|^2 |T_i|^2 +\!
2\left| \bra{\bfe_i }D_i \ket{\bfz}+{1\over\sqrt{\bb_i}}
\right|^2 |T_j|^2 ,\label{eq:vij}
\EE
and
\BE
V_{i\B{j}}(T_i ,T_j )=
4|T_i|^2 |T_j|^2.
\EE

We can integrate out the $D$ fields from (\ref{eq:tachyonpotential}) and
get the effective potential $V_{\rm eff}(\bfT)$ for the scalar field
$T_i$~\cite{AGMS}. Namely we set $D_i$ to their stationary values and search
for the minimum of $V(\bfT,\bfD)$. Because the elements
$\bra{\bfn+\bfe_i}D_i\ket{\bfn}$ with $\bfn\neq\bfz$ appear only in
$V_{i\B{i}}$, we should minimize $V_{i\B{i}}$ with respect to them and get
\BE
\left|\bra{\bfn+\bfe_i }D_i \ket{\bfn}
+\sqrt{n_i +1\over\bb_i}\right|^2 =
\left|
\bra{\bfn}D_i \ket{\bfn-\bfe_i }+\sqrt{n_i\over\bb_i}\right|^2
+{1\over\bb_i}, \label{eq:dneq0}
\EE
for $\bfn\neq\bfz$. By eliminating $\bra{\bfn+\bfe_i }D_i \ket{\bfn}, \;
(\bfn\neq\bfz)$ by (\ref{eq:dneq0}), we obtain
\BEA
V(\bfT,\bra{\bfe_i}D_i\ket{\bfz})&=&
\sum_{i=1}^{p/2}\left(|T_i|^2 -{1\over\bb_i}\right)^2
+\sum_{i=1}^{p/2}\left(|T_i|^2 -
\left|\bra{\bfe_i}D_i\ket{\bfz}+{1\over\sqrt{\bb_i}}\right|^2
+{1\over\bb_i}\right)^2 \NN\\
&&+2\sum_{i\neq j}
\left| \bra{\bfe_j }D_j \ket{\bfz}+{1\over\sqrt{\bb_j}}
\right|^2 |T_i|^2+2\sum_{i\neq j}|T_i|^2 |T_j|^2.
\label{eq:pot0}
\EEA
Let us eliminate $\bra{\bfe_i}D_i\ket{\bfz}$ in (\ref{eq:pot0}) and
obtain $V_{\rm eff}(\bfT)$. According to the area to which $T_i$ belong,
the values of $\bra{\bfe_i}D_i\ket{\bfz}$ that minimize (\ref{eq:pot0}) are
determined as follows:
\begin{enumerate}
\item If $T_i$ satisfy
$|T_i|^2 -\sum_{i\neq j}|T_j|^2+1/\bb_i\geq 0$,
\BE
\left|\bra{\bfe_i}D_i\ket{\bfz}+1/\sqrt{\bb_i}\right|^2=
|T_i|^2 -\sum_{i\neq j}|T_j|^2+1/\bb_i.
\EE
\item If $|T_i|^2 -\sum_{i\neq j}|T_j|^2+1/\bb_i< 0$,
\BE
\left|\bra{\bfe_i}D_i\ket{\bfz}+1/\sqrt{\bb_i}\right|^2=0.
\EE
\end{enumerate}

With the above preparations, we can obtain $V_{\rm eff}(\bfT)$ in each case.\\
$\bullet$ \underline{\it D0-D2}\\
Because $\bb_1\geq 0$ and we have only $T_1$, we obtain~\cite{AGMS}
\BE
V_{\rm eff}(T_1)=\left(|T_1|^2-{1\over\bb_1}\right)^2,
\EE
which is minimized at $|T_1|=1/\sqrt{\bb_1}$ to give $V_{\rm min} = 0$.
Therefore, we obtain
\BE
V_{\rm eff}(0)-V_{\rm min} = {1\over\bb_1^2}.
\EE
The minimum corresponds to pure D2-brane; the D0-brane dissolves completely
into the D2-brane.

\noindent
$\bullet$ \underline{\it D0-D4}\\
Without loss of generality, we can assume $\bb_1\geq\bb_2\geq 0$.
There are three regions of $T_i$ to be considered separately:
\begin{description}
\item[Region I:] $-1/\bb_1\leq |T_1|^2-|T_2|^2\leq 1/\bb_2$
\BEA
V_{\rm eff}(T_1,T_2)&=&
8|T_1|^2 |T_2|^2
+2\left({1\over\bb_2}-{1\over\bb_1}\right)|T_1|^2
+2\left({1\over\bb_1}-{1\over\bb_2}\right)|T_2|^2
+{1\over\bb_1^2}+{1\over\bb_2^2}.\NN\\
\label{eq:region1}
\EEA
\item[Region II:] $|T_1|^2-|T_2|^2 <-1/\bb_1$
\BE
V_{\rm eff}(T_1,T_2)=
|T_1|^4+6|T_1|^2|T_2|^2+|T_2|^4+{2\over\bb_2}|T_1|^2
-{2\over\bb_2}|T_2|^2+{2\over\bb_1^2}+{1\over\bb_2^2}.
\label{eq:region2}
\EE
\item[Region III:] $|T_1|^2-|T_2|^2 >1/\bb_2$
\BE
V_{\rm eff}(T_1,T_2)=
|T_1|^4+6|T_1|^2|T_2|^2+|T_2|^4-{2\over\bb_1}|T_1|^2
+{2\over\bb_1}|T_2|^2+{1\over\bb_1^2}+{2\over\bb_2^2}.
\label{eq:region3}
\EE
\end{description}
In Region I, because the coefficient of $|T_1|^2$ is positive, the local
minimum there is realized at $|T_1|^2=0$ and $|T_2|^2 = 1/\bb_1$.
In Region II, $V_{\rm eff}(\bfT)$ is a uniformly increasing function of
$|T_1|^2$. Therefore, the local minimum is at $|T_1|^2 =0$ again,
and (\ref{eq:region2}) becomes
\BE
V_{\rm eff}(0,T_2)=\left(|T_2|^2-{1\over\bb_2}\right)^2+
{2\over\bb_1^2}.
\label{eq:region2'}
\EE
In summary, if $\bb_1\geq\bb_2$, the minimum of $V_{\rm eff}(\bfT)$ is
$V_{\rm min}=2/\bb_1^2$.

The discussion in the case $\bb_1\leq\bb_2$ is similar to the above.
We thus obtain
\BE
V_{\rm eff}(0,0)-V_{\rm min}=
\left|{1\over\bb_1^2}-{1\over\bb_2^2}\right|,
\EE
which vanishes when the BPS condition (\ref{eq:d4bpsb}) is satisfied.

A couple of remarks are in order. First, as long as $\bb_1\neq\bb_2$,
the coefficient of either $|T_1|^2$ or $|T_2|^2$ in $V_{\rm eff}(\bfT)$ is
negative around $T_i=0$. Therefore, the one-soliton state $C_i^{(0)}$ is
stable only if $\bb_1=\bb_2$. Second, it is clear from the potential
\p{eq:tachyonpotential} that it has a global minimum at the trivial solution
\p{sol}, which corresponds to pure D4-brane. The reason why we do not see
this in our potential is that we have restricted the form of $D_i$ within
(\ref{eq:ansaetze}) for simplicity. The obtained minimum of
$V_{\rm eff}(\bfT)$ with $|T_1|^2=0$ and $|T_2|^2=1/\bb_1$ for $\bb_1\geq\bb_2$
is certainly one of the minima but is not a global minimum of the full
potential. This implies that the D0-brane could decay to this point but
would further decay towards D4 when the system is not BPS. Even so, it is
interesting to explicitly calculate the potential and see that it has
various branches where we find local minima. It is left for a future study
to examine whether there is a monotonously descending way to connect
$T_i=0$ and another minimum if we put off the condition (\ref{eq:ansaetze}).

\noindent
$\bullet$ \underline{\it D0-D6}\\
Suppose that $\bb_1\geq\bb_2\geq\bb_3$ and it is easy to show that the
potential has minimum at $T_1 = T_2=0$, as in the D0-D4. The potential then
becomes
\BE
V_{\rm eff}(T_3)=\left\{ \begin{array}{lc}
-|T_3|^4+2\left({1\over\bb_1}+{1\over\bb_2}-{1\over\bb_3}\right)
|T_3|^2 +{1\over\bb_1^2}+{1\over\bb_2^2}+{1\over\bb_3^2}&
(|T_3|^2\leq {1\over\bb_1})\\
2\left({1\over\bb_2}-{1\over\bb_3}\right)
|T_3|^2 +{2\over\bb_1^2}+{1\over\bb_2^2}+{1\over\bb_3^2}&
({1\over\bb_1}\leq |T_3|^2\leq {1\over\bb_2})\\
\left(|T_3|^2-{1\over\bb_3}\right)^2
+2\left({1\over\bb_1^2}+{1\over\bb_2^2}\right)&
(|T_3|^2\geq {1\over\bb_2})\\
\end{array}\right. .\label{eq:d6pot}
\EE
For $1/\bb_1+1/\bb_2-1/\bb_3>0$, the point $T_i=0$ is a local minimum
of $V_{\rm eff}(\bfT)$, so that the D0-D6 is stable. On the other hand,
if $1/\bb_1+1/\bb_2-1/\bb_3 < 0$, the vacuum with $T_1=T_2=T_3=0$ is
unstable and decays into other local minimum. One possible minimum is
the one from the potential \p{eq:d6pot}, with
\BE
V_{\rm eff}(\bfz)-V_{\rm min}=
-{1\over\bb_1^2}-{1\over\bb_2^2}+{1\over\bb_3^2}.
\label{minimum06}
\EE
There also exists a minimum in the original potential \p{eq:tachyonpotential}
at the trivial solution~\p{sol}. So if this D0-D6 is unstable, we again
expect that it may decay into the above solution but that would further
decay to pure D6.

\noindent
$\bullet$ \underline{\it D0-D8}\\
With $\bb_1\geq\bb_2\geq\bb_3\geq\bb_4$, the potential is again minimum
at $T_1=T_2=T_3=0$ and reduces to
\BE
V_{\rm eff}(T_4)=\left\{ \begin{array}{lc}
-2|T_4|^4+2\left({1\over\bb_1}+{1\over\bb_2}+{1\over\bb_3}-{1\over\bb_4}\right)
|T_4|^2 +{1\over\bb_1^2}+{1\over\bb_2^2}+{1\over\bb_3^2}+{1\over\bb_4^2}&
(|T_4|^2\leq {1\over\bb_1})\\
-|T_4|^4+2\left({1\over\bb_2}+{1\over\bb_3}-{1\over\bb_4}\right)
|T_4|^2 +{2\over\bb_1^2}+{1\over\bb_2^2}+{1\over\bb_3^2}+{1\over\bb_4^2}&
({1\over\bb_1}\leq |T_4|^2\leq {1\over\bb_2})\\
2\left({1\over\bb_3}-{1\over\bb_4}\right)
|T_4|^2 +{2\over\bb_1^2}+{2\over\bb_2^2}+{1\over\bb_3^2}+{1\over\bb_4^2}&
({1\over\bb_2}\leq |T_4|^2\leq {1\over\bb_3})\\
\left(|T_4|^2-{1\over\bb_4}\right)^2
+2\left({1\over\bb_1^2}+{1\over\bb_2^2}+{1\over\bb_3^2}\right)&
(|T_4|^2\geq {1\over\bb_3})\\
\end{array}\right. .\label{eq:d8pot}
\EE
For $1/\bb_1+1/\bb_2+1/\bb_3-1/\bb_4>0$ close to the new supersymmetric
locus~\p{eq:d8-+++1}, the point $T_i=0$ is a local minimum of
$V_{\rm eff}(\bfT)$, so that the D0-D6 is stable. On the other hand, if
$1/\bb_1+1/\bb_2+1/\bb_3-1/\bb_4< 0$, the vacuum with $T_1=T_2=T_3=T_4=0$
is unstable and decays into other local minima. The analysis is then similar
to the D0-D6 case.

At the supersymmetric loci~(\ref{eq:d8--++0}), the system is always stable.

In the course of writing this paper, ref.~\cite{R} came to our attention
where the spectrum of the D0-D4 system is discussed. A brief discussion
of other cases is also given. We thank the author for pointing this out.

\section*{Acknowledgements}

YI thanks K. Furuuchi for enlightening discussions.
NO would like to thank P.K. Townsend for valuable correspondence.
AF was partially supported by The Yukawa Memorial Foundation.
The work of NO was supported in part by Grants-in-Aid for Scientific
Research Nos. 99020, 12640270 and Grant-in-Aid on the Priority Area:
Supersymmetry and Unified Theory of Elementary Particles.
\\

\noindent
{\bf\Large Appendix}

\appendix

\section{Mass matrix and mass eigenvalues on D0-D$p$}
\label{sec:A}
\setcounter{equation}{0}

By substituting (\ref{fluc}) to (\ref{action}), and evaluating it up to
the second order in the fluctuations, we obtain the mass matrix.
For this sake, we define
\BE
W_i ({\bf n})=\bra{{\bf 0}}W_i \Sd\ket{{\bf n}},\quad
T_i ({\bf n})=\bra{{\bf 0}}T_i \Sd\ket{{\bf n}}.
\EE
Let ${\bf e}_j$ be $(p/2)$-dimensional unit vectors.
The contribution from each term is as follows:
\begin{itemize}
\item quadratic terms in ${\rm Tr}\,\frac{1}{2}([C_i , \bC_i ]+
\frac{1}{\bb_i})^2$:
\begin{eqnarray}
&&
\frac{1}{\bb_i}\mbox{Tr}(W_i\bW_i-T_i\bT_i )
+\mbox{Tr}
(W_i\bCz_i -T_i \Cz_i)(\Cz_i\bW_i -\bCz_i\bT_i)\nonumber\\
&=&
\sum_{n_1=0}^\infty\cdots\sum_{n_{p/2}=0}^\infty
\Biggl\{\frac{1}{\bb_i}|W_i({\bf n})|^2
-\frac{1}{\bb_i}|T_i({\bf n})|^2 \NN\\
&&\hs{30}
+\biggl|\sqrt{\frac{n_i}{\bb_i}}W_i({\bf n-e}_i)
-\sqrt{\frac{n_i+1}{\bb_i}}
T_i({\bf n+e}_i)\biggr|^2\Biggr\},\label{A1}
\end{eqnarray}
where we have dropped the terms $\frac{1}{2}(\frac{1}{\bb_i})^2$ and
$\frac{1}{2}\mbox{Tr} ([a_i^\dagger , \bD_i ]+[D_i , a_i ])^2$,
which are not necessary for the analysis of the mass spectrum.

\item quadratic terms in $\mbox{Tr}([C_i , \bC_j][C_j , \bC_i])$:
\begin{eqnarray}
&&
\mbox{Tr}\bigl(
(W_i\bCz_j-T_j\Cz_i)(\Cz_j\bW_i-\bCz_i\bT_j)
+(\Cz_i\bW_j-\bCz_j\bT_i)(W_j\bCz_i-T_i\bCz_j)\bigr)\nonumber\\
&=&\sum_{n_1=0}^\infty\cdots\sum_{n_{p/2}=0}^\infty
\Biggl\{\biggl|\sqrt{\frac{n_j}{\bb_j}}W_i({\bf n-e}_j)
-\sqrt{\frac{n_i+1}{\bb_i}}T_j({\bf n+e}_i)\biggr|^2\NN\\
&&\hs{40}
+\biggl|\sqrt{\frac{n_i}{\bb_i}}W_j({\bf n-e}_i)
-\sqrt{\frac{n_j+1}{\bb_j}}T_i({\bf n+e}_j)\biggr|^2
\Biggr\}.\label{A2}
\end{eqnarray}

\item quadratic terms in $\mbox{Tr}([C_i , C_j][\bC_j , \bC_i])$:
\begin{eqnarray}
&&\mbox{Tr}\bigl(
(W_i\bCz_j-W_j\bCz_i)(\bCz_j\bW_i-\bCz_i\bW_j)
+(\Cz_i\bT_j-\Cz_j\bT_i)(T_j\bCz_i-T_i\bCz_j)\bigr)\NN\\
&=&\sum_{n_1=0}^\infty\cdots\sum_{n_{p/2}=0}^\infty
\Biggl\{\biggl|\sqrt{\frac{n_j+1}{\bb_j}}W_i({\bf n+e}_j)
-\sqrt{\frac{n_i+1}{\bb_i}}W_j({\bf n+e}_i)\biggr|^2
\NN\\
&&\hs{40}+\biggl|\sqrt{\frac{n_i}{\bb_i}}T_j({\bf n-e}_i)
-\sqrt{\frac{n_j}{\bb_j}}T_i({\bf n-e}_j)\biggr|^2
\Biggr\}. \label{A3}
\end{eqnarray}
\end{itemize}

By assembling the terms (\ref{A1})-(\ref{A3}), we obtain the quadratic
terms of the fluctuations in the Lagrangian.
Defining a $p$-dimensional column vector
\BE
{\cal V}(\bfn)=
\left(
W_1({\bf n-e}_1),\cdots,W_{p/2}({\bf n-e}_{p/2}),
T_1({\bf n+e}_1),\cdots,T_{p/2}({\bf n+e}_{p/2})
\right)^t
\label{eq:vector}
\EE
and a $p\times p$ symmetric matrix, whose upper diagonal
elements are given by
\BE
\ssl{\cal M}(\bfn)\ssr_{ij}=
\left\{ \begin{array}{cl}
{n_i+1\over\bb_i}+\sum_{k\neq i}^{p/2}{n_k+1\over\bb_k}&\quad
(1\leq i=j\leq p/2)\\
{n_{i-p/2}\over\bb_{i-p/2}}
+\sum_{k\neq i-p/2}^{p/2}{2n_k+1\over\bb_k}&\quad
(p/2+1\leq i=j \leq p)\\
-\sqrt{n_i n_j\over\bb_i\bb_j}&\quad
(1\leq i<j\leq p/2)\\
-\sqrt{(n_{i-p/2}+1)(n_{j-p/2}+1)\over
            \bb_{i-p/2}\bb_{j-p/2}}&\quad
(p/2+1\leq i<j\leq p)\\
-\sqrt{n_i(n_{j-p/2}+1)\over\bb_i\bb_{j-p/2}}&\quad
(1\leq i\leq p/2,\,p/2+1\leq j\leq p)
\end{array}\right. ,
\label{eq:matrix}
\EE
we can see that the sum of the quadratic terms are written as
\begin{equation}
\label{quadra}
  (\mbox{sum of the quadratic terms})=
\sum_{n_1,\cdots,n_{p/2}=-1}^\infty
{\cal V}({\bf n})^\dagger{\cal M}(\bfn){\cal V}({\bf n}).
\end{equation}
Here and in what follows, we should set $W_i ({\bf m})=T_i ({\bf m})=0$
if $m_j <0$ for some $j$.

Let us diagonalize the mass matrix ${\cal M}(\bfn)$ to obtain the mass
eigenvalues. Before going into the details, we should note here that
from the Gauss law constraint (\ref{Gauss}) the quantity
\BE
\sum_{i=1}^{p/2}\left(
\sqrt{n_i\over\bb_i}W_i({\bf n}-{\bf e}_i)+
\sqrt{n_i+1\over\bb_i}T_i({\bf n}+{\bf e}_i)
\right)
\label{eq:Gc}
\EE
is time-independent and thus does not propagate.
This fact is important in the following discussions.

\subsection{D0-D2}

In D0-D2~\cite{AGMS}, the quadratic terms are
\begin{eqnarray}
&&-\frac{1}{\bb_1}|T_1(0)|^2+0\cdot|T_1(1)|^2\NN\\
&&\hs{10}+\sum_{n_1=1}^\infty
(\bW_1 (n_1-1),\bT_1(n_1+1))
\left(
\begin{array}{cc}
\frac{n_1+1}{\bb_1} & -\frac{\sqrt{(n_1+1)n_1}}{\bb_1}\\
-\frac{\sqrt{(n_1+1)n_1}}{\bb_1} & \frac{n_1}{\bb_1}
\end{array}
\right)
\left(
\begin{array}{c}
W_1(n_1-1)\\
T_1(n_1+1)
\end{array}
\right)\NN\\
&=&-\frac{1}{\bb_1}|T_1(0)|^2+0\cdot|T_1(1)|^2\NN\\
&&+\sum_{n_1=1}^{\infty} 0\cdot \left| \sqrt{n_1\over 2n_1+1}W_1(n_1-1)
+\sqrt{n_1+1\over 2n_1+1}T_1(n_1+1) \right|^2 \NN\\
&&+\sum_{n_1=1}^{\infty}{2n_1+1\over\bb_1} \left|
\sqrt{n_1+1\over 2n_1+1}W_1(n_1-1)-\sqrt{n_1\over 2n_1+1}T_1(n_1+1)
\right|^2. \label{B1D2}
\end{eqnarray}
Though the second and third terms in (\ref{B1D2}) contain zero eigenvalues,
they correspond to the unphysical modes (\ref{eq:Gc}).
Therefore, the physical mass spectrum is $(2n_1 +1)/\bb_1$
($n_1 = -1,1,2, \cdots$) without any degeneracy. Note that $n_1=0$ is
absent in this spectrum.

\subsection{D0-D4}
In D0-D4, the quadratic terms are
\begin{eqnarray}
&&\biggl(-\frac{1}{\bb_1}+\frac{1}{\bb_2}\biggr)|T_1(0,0)|^2+
\biggl(\frac{1}{\bb_1}-\frac{1}{\bb_2}\biggr)|T_2(0,0)|^2\NN\\
&&+\left(
  \begin{array}{c}
T_1(1,0)\\
T_2(0,1)
  \end{array}
\right)^\dagger
\left(
  \begin{array}{cc}
\frac{1}{\bb_2} & -\sqrt{\frac{1}{\bb_1\bb_2}}\\
 -\sqrt{\frac{1}{\bb_1\bb_2}} & \frac{1}{\bb_1}
\end{array}
\right)
\left(
  \begin{array}{c}
T_1(1,0)\\
T_2(0,1)
  \end{array}
\right)\NN\\
&&
+\sum_{n_2=1}^\infty\biggl(-\frac{1}{\bb_1}
+\frac{2n_2+1}{\bb_2}\biggr)|T_1(0,n_1)|^2
+\sum_{n_2=1}^\infty\biggl(\frac{2n_1+1}{\bb_1}
-\frac{1}{\bb_2}\biggr)|T_2(n_1,0)|^2\NN\\
&&+\sum_{n_1=1}^\infty\left(
  \begin{array}{c}
W_1(n_1-1,0)\\
T_1(n_1+1,0)\\
T_2(n_1,1)
  \end{array}
\right)^\dagger
\left(
    \begin{array}{ccc}
\frac{n_1+1}{\bb_1}+\frac{1}{\bb_2} &
-\frac{\sqrt{n_1(n_1+1)}}{\bb_1} &
-\sqrt{\frac{n_1}{\bb_1\bb_2}} \\
-\frac{\sqrt{n_1(n_1+1)}}{\bb_1} &
\frac{n_1}{\bb_1}+\frac{1}{\bb_2} &
-\sqrt{\frac{n_1+1}{\bb_1\bb_2}} \\
-\sqrt{\frac{n_1}{\bb_1\bb_2}} &
-\sqrt{\frac{n_1+1}{\bb_1\bb_2}} &
\frac{2n_1+1}{\bb_1}
    \end{array}\right)
\left(
  \begin{array}{c}
W_1(n_1-1,0)\\
T_1(n_1+1,0)\\
T_2(n_1,1)
  \end{array}
\right)\NN\\
&&+\sum_{n_2=1}^\infty\left(
  \begin{array}{c}
W_2(0,n_2-1)\\
T_1(1,n_2)\\
T_2(0,n_2+1)
  \end{array}
\right)^\dagger
  \left(
    \begin{array}{ccc}
\frac{1}{\bb_1}+\frac{n_2+1}{\bb_2} &
-\sqrt{\frac{n_2}{\bb_1\bb_2}} &
-\frac{\sqrt{n_2(n_2+1)}}{\bb_2}\\
-\sqrt{\frac{n_2}{\bb_1\bb_2}} &
\frac{2n_2+1}{\bb_2} &
-\sqrt{\frac{(n_2+1)}{\bb_1\bb_2}}\\
-\frac{\sqrt{n_2(n_2+1)}}{\bb_2}  &
-\sqrt{\frac{(n_2+1)}{\bb_1\bb_2}} &
\frac{1}{\bb_1}+\frac{n_2}{\bb_2}
    \end{array}\right)
\left(
  \begin{array}{c}
W_2(0,n_2-1)\\
T_1(1,n_2)\\
T_2(0,n_2+1)
   \end{array}
\right)\NN\\
&&+\sum_{n_1,n_2=1}^\infty
{\cal V}(\bfn)^\dagger {\cal M}(\bfn){\cal V}(\bfn),
\label{B1D4}
\end{eqnarray}
where ${\cal V}$ and ${\cal M}$ are defined in (\ref{eq:vector}) and
(\ref{eq:matrix}) with $p=4$, respectively. We again find that the above
matrices include zero eigenvalues. However, just as in the D0-D2 case,
we can show that the zero eigenvalues correspond to the unphysical gauge modes
(\ref{eq:Gc}). This is also true for D0-D6 and D0-D8, and we will not
comment on the zero eigenvalues in what follows.

Omitting these unphysical zero eigenvalues, the physical mass spectrum is
\BE
{2n_1+1\over\bb_1}+{2n_2+1\over\bb_1},
\EE
where $n_1,n_2=-1,0,1,2,\cdots$ except the case in which $n_1$ and
$n_2$ are $-1$ simultaneously. The degeneracy of the mass eigenvalues varies
depending on the size of mass matrices in \p{B1D4}. It is summarized in
table~1, where ``$\ast$'' means any positive integer.
\begin{center}
\begin{tabular}{|l|l|c|}
\hline
$(n_1,n_2)$&mass eigenvalue&degeneracy\\
\hline\hline
$(-1,0)$&$-1/\bb_1+1/\bb_2$&$1$\\
$(0,-1)$&$1/\bb_1-1/\bb_2$&$1$\\
$(0,0)$&$1/\bb_1+1/\bb_2$&$1$\\
$(-1,\ast)$&$-1/\bb_1+(2n_2+1)/\bb_2$&$1$\\
$(\ast,-1)$&$(2n_1+1)/\bb_1-1/\bb_2$&$1$\\
$(0,\ast)$&$1/\bb_1+(2n_2+1)/\bb_2$&$2$\\
$(\ast,0)$&$(2n_1+1)/\bb_1+1/\bb_2$&$2$\\
$(\ast,\ast)$&$(2n_1+1)/\bb_1+(2n_2+1)/\bb_2$&$3$\\
\hline
\end{tabular}
\vs{1}

Table 1: Degeneracy of mass eigenvalues in D0-D4
\end{center}

\subsection{D0-D6}

Here (and for the following D0-D8) we do not give the explicit mass matrices
but simply present the mass eigenvalues:
\BE
{2n_1+1\over\bb_1}+{2n_2+1\over\bb_2}+{2n_3+1\over\bb_3},
\quad n_i =-1,0,1,2,\cdots,
\EE
where any two of $n_i$ cannot be simultaneously $-1$.

The degeneracies are shown in table~2. The eigenvalues are symmetric under
the permutation of $i$, so the degeneracy, for example, of the eigenvalues
with $(n_1,n_3,n_2)$, is the same as that with $(n_1,n_2,n_3)$.
Therefore, we show only those for $n_1\leq n_2\leq n_3$.

\begin{center}
\begin{tabular}{|l|l|c|}
\hline
$(n_1,n_2,n_3)$&mass eigenvalue&degeneracy\\
\hline\hline
$(-1,0,0)$&$-1/\bb_1+1/\bb_2+1/\bb_3$&$1$\\
$(-1,0,\ast)$&$-1/\bb_1+1/\bb_2+(2n_3+1)/\bb_3$&$1$\\
$(-1,\ast,\ast)$&$-1/\bb_1+(2n_2+1)/\bb_2+(2n_3+1)/\bb_3$&$1$\\
$(0,0,0)$&$1/\bb_1+1/\bb_2+1/\bb_3$&$2$\\
$(0,0,\ast)$&$1/\bb_1+1/\bb_2+(2n_3+1)/\bb_3$&$3$\\
$(0,\ast,\ast)$&
$1/\bb_1+(2n_2+1)/\bb_2+(2n_3+1)/\bb_3$&$4$\\
$(\ast,\ast,\ast)$&
$(2n_1+1)/\bb_1+(2n_2+1)/\bb_2+(2n_3+1)/\bb_3$&$5$\\
\hline
\end{tabular}
\vs{1}

Table 2: Degeneracy of mass eigenvalues in D0-D6
\end{center}

\subsection{D0-D8}

The mass eigenvalues are
\BE
{2n_1+1\over\bb_1}+{2n_2+1\over\bb_2}
+{2n_3+1\over\bb_3}+{2n_4+1\over\bb_4},
\quad n_i =-1,0,1,2,\cdots,
\EE
where any two of $n_i$ cannot be simultaneously $-1$.
The degeneracies with $n_1\leq n_2\leq n_3\leq n_4$ are given in table~3.
\begin{center}
\begin{tabular}{|l|l|c|}
\hline
$(n_1,n_2,n_3,n_4)$&mass eigenvalue&degeneracy\\
\hline\hline
$(-1,0,0,0)$&
$-1/\bb_1+1/\bb_2+1/\bb_3+1/\bb_4$&$1$\\
$(-1,0,0,\ast)$&
$-1/\bb_1+1/\bb_2+1/\bb_3+(2n_4+1)/\bb_4$&$1$\\
$(-1,0,\ast,\ast)$&
$-1/\bb_1+1/\bb_2+(2n_3+1)/\bb_3+(2n_4+1)/\bb_4$&$1$\\
$(-1,\ast,\ast,\ast)$&
$-1/\bb_1+(2n_2+1)/\bb_2+(2n_3+1)/\bb_3+(2n_4+1)/\bb_4$&$1$\\
$(0,0,0,0)$&
$1/\bb_1+1/\bb_2+1/\bb_3+1/\bb_4$&$3$\\
$(0,0,0,\ast)$&
$-1/\bb_1+1/\bb_2+1/\bb_3+(2n_4+1)/\bb_4$&$4$\\
$(0,0,\ast,\ast)$&
$1/\bb_1+1/\bb_2+(2n_3+1)/\bb_3+(2n_4+1)/\bb_4$&$5$\\
$(0,\ast,\ast,\ast)$&
$-1/\bb_1+(2n_2+1)/\bb_2+(2n_3+1)/\bb_3+(2n_4+1)/\bb_4$&$6$\\
$(\ast,\ast,\ast,\ast)$&
$(2n_1+1)/\bb_1+(2n_2+1)/\bb_2+(2n_3+1)/\bb_3+(2n_4+1)/\bb_4$
&$7$\\
\hline
\end{tabular}\\
\vs{1}
~Table 3: Degeneracy of mass eigenvalues in D0-D8
\end{center}

\newcommand{\ANN}[3]{Ann. Phys. (NY) {\bf #1} {(#2)} {#3}}
\newcommand{\CMP}[3]{Comm. Math. Phys. {\bf #1} {(#2)} {#3}}
\newcommand{\IJMP}[3]{Int. J. Mod. Phys. {\bf #1} {(#2)} {#3}}
\newcommand{\JHEP}[3]{JHEP {\bf #1} {(#2)} {#3}}
\newcommand{\NP}[3]{Nucl. Phys. {\bf #1} {(#2)} {#3}}
\newcommand{\PL}[3]{Phys. Lett. {\bf #1} {(#2)} {#3}}
\newcommand{\PR}[3]{Phys. Rev. {\bf #1} {(#2)} {#3}}
\newcommand{\PRL}[3]{Phys. Rev. Lett. {\bf #1} {(#2)} {#3}}
\newcommand{\PTP}[3]{Prog. Theor. Phys. {\bf #1} {(#2)} {#3}}
\newcommand{\NULL}[3]{ {\bf #1} {(#2)} {#3}}
\newcommand{\ibid}[3]{{\it ibid.} {\bf #1} {(#2)} {#3}}

\end{document}